\documentclass{camera}
\begin{document}

\title{THE CLIC STUDY:\\
TAKING THE LEP LEGACY\\ 
TO THE MULTI-TEV FRONTIER}

\author{Marco Battaglia}
\organization{\vspace*{-2.5cm} CERN - EP, CH-1211 Geneva 23 (Switzerland)}
\maketitle


\vspace*{-2.0cm}

With the completion of the {\sc Lep-2} operation, the exploration of the high 
energy frontier in $e^+e^-$ physics is most likely to be inherited by a 
linear collider (LC). In fact, the $E_{beam}^4$ scaling of the beam energy 
loss in circular storage rings makes them no longer a viable choice for 
collision energies beyond 
about 300~GeV.  A rich harvest of physics results should become accessible at 
centre-of-mass energies $\sqrt{s} \simeq$ 500~GeV, that can be obtained with 
super-conducting RF structures, as proposed  by the {\sc Tesla} project~\cite{tesla_tdr},
or also with X-band warm cavities~\cite{xband}. 
If the electro-weak symmetry breaking is realised through the Higgs mechanism, 
whose onset is manifested by the existence of an elementary Higgs boson, a 500~GeV 
LC at high luminosity will promote Higgs physics into the domain of precision 
measurements. In the study of the Higgs sector, the LC may thus become to 
the {\sc Lhc},  the equivalent of what {\sc Lep} has been to the 
{\sc $Sp\bar{p}S$}, in the study of the $Z^0$ and $W^{\pm}$ bosons.

If the Higgs field is indeed responsible for the masses of the $Z^0$ 
and $W^{\pm}$ bosons, that {\sc Lep} has measured so accurately, as well as for
those of the fermions, a further exploration of the TeV frontier will become 
necessary to fully probe its structure and nature and to understand the 
mechanism that stabilises the electro-weak scale giving $v \simeq$ 250~GeV 
$\simeq 10^{-17} \times M_{Planck}$. 
Furthermore, in order to precisely test the nature of signals that may be 
observed at the {\sc Lhc}, by the end of this decade, and to extend the probe 
for new physics beyond its reach, a lepton collider 
able to deliver collisions at energies in the multi-TeV range will be 
required. This motivates the development of new techniques of particle 
acceleration, beyond those presently considered for a TeV-class LC.

The {\sc Clic} project, carried out at {\sc Cern} since 1987, aims at 
developing and validating the two beam acceleration scheme to provide 
$e^+e^-$ beams colliding at $\sqrt{s}$ energies from 0.5~TeV, or lower energies if
the need would arise, up to 5~TeV~\cite{clic}.  
While promising to open a new domain for experimentation 
at $e^+e^-$ colliders, {\sc Clic} also presents new challenges compared to any
other $e^+e^-$ collider project, due to its different regime of operation.
In order to balance the $1/s$ fall of s-channel cross sections, experimentation 
at multi-TeV energies requires a luminosity of the order of 
$10^{35}$cm$^{-2}$s$^{-1}$. In the {\sc Clic} scheme, the necessary RF power 
is generated at 30~GHz, by means of a high intensity, low energy drive beam, extracted
by copper transfer structures and transferred to the main beam .
The 30~GHz choice allows to operate with a larger gradient, thus minimising
the linac total length and therefore its cost. However, since at 30~GHz the 
cavity aperture is only $\sim$4~mm, the wake-field generation becomes more 
important, compared to lower frequencies, and the beam emittance needs to be
preserved by a careful choice of the parameters and by enforcing tighter 
alignment tolerances. At the same time, at $\sqrt{s}$ = 3-5~TeV, 
$e^+e^-$ collisions take place in a regime of large 
beamstrahlung ($\Upsilon >> 1$) where the luminosity is expected to increase with the RF 
frequency and to be independent of the accelerating gradient~\cite{scaling}. In this
regime the luminosity spectrum is significantly broadened by the energy loss suffered
by the electron and positrons in the intense field generated by the incoming beam.

The design luminosity is achieved with a bunch charge, corresponding to only
one tenth of that accelerated at {\sc Lep} but focused to a spot four orders 
of magnitude smaller, in the horizontal and vertical planes, at the 
interaction point.

The definition of the {\sc Clic} physics programme still requires essential 
data that is likely to become available only after the first years of 
{\sc Lhc} operation and, possibly, also the results from a LC operating 
at lower energies~\cite{signatures}. At present we thus have to envisage several 
possible scenarios for the fundamental questions to be addressed by HEP experiments 
in the second decade of this new century. 

If new particles will have been observed, either at the {\sc Lhc} or in lower energy 
$e^+e^-$ data, {\sc Clic} has the potential to complement the probe to their nature.
In the case of an elementary Higgs boson, even after the {\sc Lhc} and a 500~GeV LC will 
have studied in details its properties, the ultimate test that the observed boson 
is the manifestation of the scalar potential responsible for electro-weak symmetry 
breaking will come from the determination of 
the triple and, possibly, the quartic self-couplings as can be obtained at a high 
luminosity multi-TeV $e^+e^-$ collider.
If Supersymmetry is realised in Nature, signals for supersymmetric partners of the 
known particles should be observed at both the {\sc Lhc} and in lower energy
$e^+e^-$ collisions. Beyond discovery, it will be crucial to accurately study the 
properties of all these particles (masses, couplings and quantum numbers), to understand
the underlying model of symmetry breaking and to determine the theory parameters. This 
will likely require $e^+e^-$ data at $\sqrt{s}$ energies in excess to 1~TeV.

A striking manifestation of new physics in the multi-TeV region will 
come from the sudden increase of the $e^+e^- \rightarrow f \bar f$ cross 
section signalling the s-channel production of a new particle. There are several 
scenarios predicting new resonances in the mass range of interests for {\sc Clic}. 
A first exemplificative category of New Physics, yielding such signatures is represented
by additional gauge bosons like a $Z'$. These are common to both Grand Unification 
inspired $E_6$ models and to Left-Right symmetric models. Further, Kaluza-Klein 
excitations of the graviton, of $Z$ and $\gamma$ gauge bosons, or both, are a 
distinctive feature of models of quantum gravity with extra-spatial dimensions. 
Finally, models with strong symmetry breaking may be characterised by the appearance 
of resonances at the TeV scale in $WW$ scattering. If such new states would become 
directly observable at {\sc Clic}, the study of electroweak observables around the 
resonance will be required to precisely determine their nature and
couplings. Even if no new particle will be directly produced, the comparison of 
precision electro-weak data to the Standard Model (SM) predictions has the potential to 
open a window that extends the linear collider sensitivity to new physics to scales far 
beyond the centre-of-mass energies. This study would essentially reproduce the {\sc Lep}
physics programme at energies beyond the TeV frontier and will significantly 
profit of the availability of polarised beams to enhance the sensitivity to 
deviations from the SM couplings. In particular {\sc Clic} could probe the existence of 
new gauge bosons, of extra dimensions and of other new phenomena in terms of contact 
interactions up to scales of order 30~TeV to 200~TeV. This sensitivity will investigate 
a new mass scale, well beyond that explored at the {\sc Lhc} and will require to compute
SM predictions for electro-weak observables with a comparable accuracy to that 
achieved at {\sc Lep} energies. 

While considering experimentation at a multi-TeV collider, it is also
important to verify to which extent extrapolations of the experimental 
techniques, successfully developed at {\sc Lep} and being refined in the
studies for the {\sc Tesla} project, are still applicable. This has major
consequences on the requirements for the experimental conditions at the 
{\sc Clic} interaction region and for the definition of the {\sc Clic} 
physics potential. Now, there are two main issues relevant to the applicability of 
techniques developed at {\sc Lep}/{\sc Slc} in the multi-TeV regime. These are: i) the 
increased boost of hadronic jets and ii) the large accelerator induced backgrounds, 
mostly from $\gamma\gamma \rightarrow {\mathrm{hadrons}}$ and pairs and the broad 
luminosity spectrum. 
The {\sc Lep} experience has taught us that the reconstruction of 
multi-fermion final states is best achieved by combining the independent 
response of the tracking and calorimetric detectors using an energy flow
algorithm while the parton flavour can be efficiently identified using the precise
determination of particle trajectories in the vertex detector.
As the jet collimation increases, the di-jet separation decreases and the heavy hadron
decay distance is enhanced by the large boost at multi-TeV energies, this reconstruction
techniques will need to be re-optimised in particular in few parton final states. 

\vspace{0.25cm}

The physics potential of a multi-TeV linear collider still needs to be explored in full
details. However, a first recognition of the main anticipated physics signatures shows
that many of the concepts and techniques developed at {\sc Lep} are likely to still 
provide a direction for probing the next energy frontier beyond the {\sc Lhc}.

%
\end{document}